\documentstyle[a4,epsf]{article}

\begin{document}

\begin{center}
{\huge\bf Quantization Of Cyclotron Motion and Quantum Hall Effect}
\end{center}

\vspace{1cm}
\begin{center}
{\large\bf 
F.GHABOUSSI}\\
\end{center}

\begin{center}
\begin{minipage}{8cm}
Department of Physics, University of Konstanz\\
P.O. Box 5560, D 78434 Konstanz, Germany\\
E-mail: ghabousi@kaluza.physik.uni-konstanz.de
\end{minipage}
\end{center}

\vspace{1cm}

\begin{center}
{\large{\bf Abstract}}
\end{center}

\begin{center}
\begin{minipage}{12cm}
We present a two dimensional model of IQHE in accord with the cyclotron motion. The quantum equation of the QHE curve and a new definition of filling factor are also given. 
\end{minipage}
\end{center}

\newpage
It is known that QHE is a two dimensional topological invariant quantum effect which is caused by a strong magnetic field on two dimensional electronic samples. Thus in the usual three dimensional samples only the classical Hall effect can be observed. Therefore the two dimensionality of sample is essential for the {\it quantum} Hall effect. Recall that the QHE is related, as a two dimensional {\it topological invariant} effect, with topological invariants of the two dimensional manifold of QHE sample; and that the main topological invariant of a two dimensional manifold is its area with $(L^2 )$ measure. Therefore the QHE is related with an invariant $(L^2 )$ measure. On the other hand as a topological {\it quantum} effect, the QHE is related with a {\it quantization} of such a topological measure. Furthermore, since QHE appears on two dimensional {\it electronic} samples, therefore the quantum property of QHE is due to the quantum property of electrons under the QHE conditions. In this sense the QHE is related with a quantized measure of area in units of some square of length $l^2$, which is given by a quantum measure of square of length for electrons on the QHE sample. Recall that such a quantum measure of square of length for electrons, is given by the square of magnetic length: $l^2 _B := \displaystyle{\frac{\hbar}{e B}}$, where $B$ is the constant magnetic field which acts on electrons. Therefore QHE as a two dimensional topological invariant quantum effect is related with the two dimenaional topological invariant quantum of area, i. e. with $l^2 _B := \displaystyle{\frac{\hbar}{e B}}$, which is a {\it quantum invariant}.
To recognize these aspects of QHE phenomenologically, one should consider the importent role played by the cyclotron radius or the magnetic length in the phenomenology of the cyclotron motion and QHE \cite{aoki}. Neverthelss a square of length or area $( \sim l_B ^2 )$ is {\it only} a topological invariant of a two dimensional manifold. Therefore, since the phenomenological description of magnetic quantization or cyclotron motion of electrons in QHE results in a two dimensional quantum invariant $l^2 _B$. Thus the theory of QHE should be a two dimensional theory, i. e. it should be defined on a two dimensional space with such a two dimensional invariant. 

In view of this anylysis of physical and topological structure of QHE, it seems that a two dimensional topological model is appropriate to describe the QHE, thus as we show, such a model can describe also the preference of edge currents in QHE \cite{kkkk}. 
 
We present here a two dimensional model for QHE where the two dimensional structure on the sample is governed by the quantum invariant of square of magnetic length. Our model is based on the three main properties of the QHE, i. e. on its two dimensionality, on the topological invariance of this effect and on the preference of edge currents in QHE \cite{kkkk}.

Note that, since the electronic motion which result from the interaction of electrons with gate potential, will cause the Hall current and Hall potential which can be measured with respect to the gate potential, thus one measures a potential always in accord with data of electrons which interact with that potential. Thus the Hall potential can be measured with respect to the electronic dynamics which is caused by the interaction of electrons with the gate potential. Nevertheless, as we show next, since the electron is a quantum particle, therefore the {\it uncertainty} of its quantities prevents an exact determination of the value of Hall potential with respect to the gate potential. Since they allow only measurments of potential values up to some potential uncertainties. It is these uncertainties of the measured values of Hall and gate potentials which results in an uncertainty relation between these potentials that is represented by the IQHE curve. 
Herefore we discuss first the {\it neccessity} of a two dimensional model for QHE from the point of view of quantum phenomenology of magnetic quantization, and then we show that there is indeed a canonically quantized two dimensional model which describes the IQHE.

To begin note that the coordinates of an electron on the two dimensional QHE sample should be given by $X_m; \ \ m,n = 1, 2$ where $X_1 = X$ and $X_2 = Y$ \cite{Bsing}. Nevertheless the position of such an electron possess, as a quantum particle, uncertainties $\Delta X_m$ in accord with quantum mechanics (QM) which are given by: $\Delta P_m \cdot \Delta X_m \geq \hbar$, where $\Delta P_m$ are the uncertainty of the momentum of electron. Furthermore recall that on the {\it two dimensional} sample manifold, also the related electromagnetic potential vector possess only {\it two components}. Therefore one can identify these {\it two components} with the gate and the Hall potentials. 

On the other hand, since the electromagnetic potential $A_m$ \cite{rel} can be measured only by interaction with a "charged test body" with a position $X_m$ \cite{mes} and since the value of such a potential on a two dimensional manifold is given by: $A_m = B \cdot X^n \epsilon_{mn} \ , \ \epsilon_{mn} = \epsilon_{nm} = -1$ \cite{nnnn}. Therefore the value of such a potential $A_m$, on the QHE sample under quantum conditions, can be determined only up to the quantum uncertainties:

\begin{equation}
\Delta A_m = B \cdot \Delta X^n \epsilon_{mn}
\end{equation}

Hence there are the following products of uncertainties in the quantum measurments of electromagnetic potentials on the QHE samples:

\begin{eqnarray}
&&\Delta A_x \cdot \Delta A_y = B^2 \cdot \Delta X \cdot \Delta Y\\
&& \Delta A_x \cdot \Delta X = \Delta A_y \cdot \Delta Y = B \cdot \Delta X \cdot \Delta Y
\end{eqnarray}

Moreover in view of the fact that any electron on the QHE sample fulfil a cyclotron motion which is defined by the commutator of operators for its relative coordinates \cite{aoki}:

\begin{equation}
[ \hat{Y} \ \ , \ \ \hat{X} ] = - i l^2 _B, \ \ l_B ^2 := \displaystyle{\frac{\hbar}{e B}} 
\end{equation}

Hence there exists a "quantum of length", i. e. the most minimal available length, for this motion which is given by the magnetic length $l_B$.
Therefore in view of the fact that such a quantum of length may not be undercut, the position uncertainties should be always larger than the quantum of length and should obey the following relations:

\begin{equation}
\Delta X_m \geq l_B \ \ , \ \ \Delta X \cdot \Delta Y \geq l_B ^2
\end{equation}

Thus using the definition of magnetic length in (2) and (3) and considering the inequalities (5), they can be rewritten by the inequalities:

\begin{eqnarray}
&& e \Delta A_x \cdot \Delta A_y \geq B \hbar\\
&& e \Delta A_x \cdot \Delta X = e \Delta A_y \cdot \Delta Y \geq \hbar
\end{eqnarray}

Note that these relations are related with each other in accord with $A_m = \epsilon_{mn} B \cdot X^n$. Thus, if the relation (6) holds, the relation (7) holds also and vice versa.
Therefore if we consider, as usual, Hall and gate potentials as $A_H = A_y$ and $A_G = A_x$ potential components on the sample, the following uncertainty relations are given for their measured values by the QHE methode on the sample:

\begin{eqnarray}
&& \Delta A_G \cdot \Delta A_H \geq \displaystyle{\frac{\hbar B}{e}}\\
&& \Delta A_G \cdot \Delta X = \Delta A_H \cdot \Delta Y \geq \displaystyle{\frac{\hbar}{e}}
\end{eqnarray}

In other words the measured Hall potential in the QHE are quantized with respect to the values of gate potential, as $\Delta A_H$ is quantized with respect to $\Delta A_G$.
Thus one should variefy relation (8) on the IQHE curve where the quantum relation between Hall and gate potentials is given in accord with their measured values \cite{k}. This is indeed the case, if $\Delta A_H$ is considered as the Hall potential difference between any two subsequent plateaus on this curve and $\Delta A_G$ as the breadth of one of these plateaus. In other words the product (8), which gives the areas of parallelograms on such plateaus: $a_i = (\Delta A_G \cdot \Delta A_H)$, should be invariant for all pregnant parallelograms on the {\it schematic} curve (see fig. 1). Thus if one considers the real QHE curve \cite{k}, then one confirms that the areas of pregnant parallelograms, i. e. those which can be drawn on the second and the third plateaus, are approximately equal.  
Moreover note that the uncertainty relation (8) is a finite and invariant relation on the sample. Therefore the stability of plateaus can be considered as a consequence of the invariance or the stability of relation (8) against the perturbations: $X_m \rightarrow X_m + \delta X_m$ or 
$A_m \rightarrow A_m + \delta A_m$, since the uncertainty relations contain only differences $\Delta X_m$ or $\Delta A_m$ and $\Delta ( \delta X_m ) = 0$.
Note also that, since the relation (8) gives the reciprocal relation between the actually measured values of quantized Hall potential $\Delta A_H$ with respect to the gate potential values $\Delta A_G$: Therefore for higher values of gate potential where the breadth of plateaus are larger: $(\Delta A_G >)$, the measured values of the quantized Hall potential, i. e. the heights between two subsequent plateaus becomes smaller: $(\Delta A_H <)$. On the other hand for smaller values of gate potential: $(\Delta A_G <)$ where the bredth of plateaus are smaller, the measured values of the quantized Hall potential, i. e. the heights between two subsequent plateaus becomes higher: $(\Delta A_H >)$. These cases describe the lower and upper parts of the real IQHE curve, respectively, where the plateau structure are not so pregnant as in the middle part of the curve. One can see such a reciprocal correspondance also in the schematic curve (see fig 1). In other words in the lower part of the real curve, in view of the higher gate voltage $( \sim \Delta A_G >> )$, the height between two subsequent plateaus becomes very small so that subsequent pleteaus {\it seems} to lie close together with vanishing height difference $( \sim \Delta A_H << )$. Whereas in the upper part, in view of lower gate voltage, the pleteaus {\it seems} to disappear $(\sim \Delta A_G << )$. Nevertheless the area of any parallelogram which can be drawn between two {\it subsequent} plateaus on the lower plateau of the IQHE curve, is always about $\displaystyle{\frac{\hbar B }{e}}$. In other words the relation (8) is valid for all such parallelograms on pleateaus, i. e. it is valid in principle for the whole IQHE curve, but it is difficult to varify this for lower and upper parts of the curve, where it is difficult to distinguish and to identify the plateaus.

It is worth mentioning that in accord with this model the discussed QHE curve which is given as in (8) by: $\Delta A_H = \displaystyle{\frac{\hbar B}{e \Delta A_G}}$, appears as the quantum version of the classical curve for the classical Hall potential which is given as a Coulomb potential by: $A_H ^{(classical)} = \displaystyle{\frac{B}{A_G}}$ or $A_H ^{(classical)}= \displaystyle{\frac{1}{Y}}$ (see fig. 1). Thus if one planes the the real QHE curve, then one obtains the classical Coulomb potential curve: $A_y \sim \displaystyle{\frac{1}{Y}}$. Note that such a planing implies both transitions: $\Delta A_G \rightarrow 0$ and $\Delta A_H \rightarrow 0$ which means just the {\it classical limit}, since the quantum uncertainties tends to zero in the classical limit. For weaker magnetic fields, where in view of $\Delta A_G \cdot \Delta A_H \sim \displaystyle{\frac{\hbar B}{e}}$ also the mentioned areas $a_i$ become smaller, the QHE curve tends to the classical potential curve. In other words the classical Coulomb curve $A_H = \displaystyle{\frac{B}{A_G}} = \displaystyle{\frac{1}{Y}}$ appears here as the classical limit of the quantum curve: $\Delta A_H = \displaystyle{\frac{\hbar B}{e \Delta A_G}} = \displaystyle{\frac{\hbar}{e \Delta Y}}$. Then the classical limit is obtained for $\hbar \rightarrow 0$ or $B \rightarrow 0$, where also the areas $a_i$ tend to zero. Thus by tending these areas to zero, the quantum curve approaches the classical potential curve as its classical limit. Hence the quantum Hall potential can be considered as the quantum limit of the classical Hall potential or Coulomb potential and vice versa. The existence of such a classical Coulomb potential limit $( A_H = \displaystyle{\frac{1}{Y}})$ for the quantum Hall potential $( \Delta A_H = \displaystyle{\frac{e \hbar}{ e^2  \Delta Y}})$ in this model, confirms the consistency of this model. 
Thus in view of the existence of a classical Hall effect, the consistency of the quantum model of QHE depends also on a consistent classical limit of this model. Thus it is importent to note that {\it only} a {\it two dimensional} quantum model of QHE contains:

1. A quantum relation (8) to describe the QHE curve. 

2. A quantum relation (9) which has a classical Coulomb potential limit (see fig. 1 ), in accord with the above discussed classical limit.  

The importance of classical Coulomb potential limit for the quantum Hall potential is essential, since it shows the consistency of our two dimensional model: Thus as we discussed above, in view of the {\it two dimensionality} of QHE, the two relevant Hall and gate potentials can be given withuout {\it loose of generality} by: $(A_H := B \cdot X \sim \displaystyle{\frac{1}{ L^2}} \cdot L \sim \displaystyle{\frac{1}{ L}})$ and $(A_G := - B \cdot Y \sim \displaystyle{\frac{1}{ L^2}} \cdot L \sim \displaystyle{\frac{1}{ L}})$. The generality of this description of potential components is only given on a two dimensional manifold where the field strength is constant by definition \cite{nnnn}, where the classical Hall and gate potentials are both of the Coulomb type which is of dimension $(\sim \displaystyle{\frac{1}{ L}})$. Note that also the quantized Hall and gate potentials, i. e. $\Delta A_H = \displaystyle{\frac{e \hbar}{e^2 \Delta Y}}$ and $\Delta A_G = \displaystyle{\frac{e \hbar}{e^2 \Delta X}}$ have the same dimension: $(\displaystyle{\frac{1}{ L}})$, in view of dimensionlessness of $\hbar$ and $e$ in the geometric units. 
Therefore the fact that the classical limit of QHE curve, i. e. of the quantum Hall potential curve, is exactly the classical Coulomb curve underlines the consistency of our analysis of QHE model as a two dimensional model.

Note that, in accord with QM, the uncertainty relations (7) corresponds to the commutators: $e [ \hat{A}_x \ \ , \ \ \hat{X} ] = e [ \hat{A}_y \ \ , \ \ \hat{Y} ] = -i \hbar$, where $\hat{A}_m$ and $\hat{X}_m$ are, respectively, the quantum operators for electromagnetic potential components and for the  coordinates of electron which interacts with such a potential under quantum conditions of cyclotron motion in QHE (see also relation (4)). Therefore, in view of such a correspondence between uncertainty relations and commutators in QM \cite{lan}, there should be a quantum model for QHE with these commutators and relations (7).

Indeed there is such a model for electrons interacting with a magnetic field which is introduced in the following:
The action functional for an interaction between the single electron and the electromagnetic potential is given, in accord with Stokes theorem, by the two dimensional topologically invariant action:

\begin{equation}
S = \displaystyle{\frac{1}{2}} (\int \int\limits_{surface} dP_m \wedge d X^m + e \int \int\limits_{surface} F_{mn} d X^m \wedge dX^n) = \displaystyle{\frac{1}{2}} (\oint\limits_{contour} P_m d X^m + e \oint\limits_{contour} A_m d X^m) \ \ ,
\end{equation} 

where $P_m$ and $X_m$ are the momentum and the position coordinates of electron as before and the {\it surface} integral is considered over the whole surface of QHE sample, whereas the {\it contour} integral is considered over the contour region of the sample. Thus the equality represents Stokes theorem for both kinetic and interaction terms of electron.

Note that the actual motion of a physical system takes place always on a polarized phase space which contains the half of phase space variables \cite{geq}, thus the action function of a system and its wave funtion are always function of half of phase space variable, beside the time parameter. The best example of such a polarized phase space with half of phase space variables is the configuration space which is known in the quantum theory under the position representation of wave function. The configuration space of QH-system is the two dimensional sample surface where the position of electrons of QHE sample are defined. Therefore the actual action function of the two dimensional QHE model can be defined on the polarized phase space of the QHE system, i. e. on the two dimensional configuration space of the system or on the two dimensional QHE sample. It is in this manner that we can write the action function (10) on the sample surface or equivalentely on the contour of QHE sample in view of the Stokes theorem; since the sample surface represents the polarized phase space of our QHE system. 

The Euler-Lagrange equations of system (10) are given by the Lorentz equations, in accord with $(dX^m = \dot{X}^m dt)$:

\begin{equation}
\dot{P}_m = e E_m - e \epsilon_{mn} \dot{X}^n \cdot B \ \ \,
\end{equation}

with $E_m := - \dot{A}_m$, which shows the consistency of the classical model.
Thus the canonical conjugate variables of this system are given by: $X_m$ and $\Pi_m = \displaystyle{\frac{1}{2}} (P_m + e A_m )$ which should not be confused with the velocity variables of electron: $V_m = \displaystyle{\frac{1}{2 M_e}} ( P_m - e A_m )$. Hence the canonical quantization of the system, which is equivalent to the Bohr-Sommerfeld quantization, is given by:

\begin{equation}
[ \hat{P}_m  \ \ , \  \hat{X}_m ] = -i \hbar \delta_{mn} \ \ , \ \ e [ \hat{A}_m  \ \ , \ \ \hat{X}_m ] = -i \hbar \delta_{mn} \ \ ,
\end{equation}

as expected \cite{new}.
These commutators correspond to the uncertainty relations $\Delta P_m \cdot \Delta X_m \geq \hbar$ and (6)-(9) or to:

\begin{equation}
e \Delta A_m \cdot \Delta X_m \geq \hbar \ \sim \ ( e \Delta A_H \cdot \Delta Y \geq \hbar \ , \ e \Delta A_G \cdot \Delta X \geq \hbar )
\end{equation}

Note that, in view of $A_m = \epsilon_{mn} B \cdot X^n$, the last commutator in (12) is equal to the commutator: $e [ \hat{A}_m  \ \ , \ \ \hat{A}_n ] = -i \hbar B \epsilon_{mn}$ or to $e [ \hat{A}_G  \ \ , \ \ \hat{A}_H ] = -i \hbar B$ which is related, in accord with QM, to the uncertainty relation (8). Thus one obtains the uncertainty relation (8) also from any uncertainty relation in (13), using $A_m = \epsilon_{mn} B \cdot X^n$. 
Therefore the accordance of empirical IQHE curve \cite{k}, with the uncertainty relation (8) of present model in view of the above discussion, shows the accordance of quantum structure of QHE with the quantum structure of present model.

Moreover, since the Bohr-Sommerfeld quantization is restricted by the contour integral (10) to the edge region in accord with $S = S_{(edge)} = S_{(contour)} = \oint\limits_{contour} \Pi dx^m = N h$. Therefore, in view of this restriction also the state functions $\Psi \propto \displaystyle{ e^{\frac{i}{\hbar} S}}$ are restricted to the edge states: $\Psi_{(edge)} \propto \displaystyle{ e^{\frac{i}{\hbar} S_{(edge)}}}$ which fulfil the equivalent quantization commutator (12) in accord with: $[ \hat{\Pi}_m  \ \ , \  \hat{X}_m ] \Psi_{(edge)} = -i \hbar \delta_{mn} \Psi_{(edge)}$. Hence in view of the restriction of state functions, which give the position probabilty of electrons on the QHE sample, to the edge states: $\Psi_{(edge)} \propto \displaystyle{ e^{\frac{i}{\hbar} S_{(edge)}}}$, the position of electrons is restricted to the edge region. This may explain the appearence of edge states and edge currents in QHE as is known from experimental results \cite{kkkk}. 

Note also that the last commutator in (12) is equivalent to the commutator (4) of the cyclotron motion, in accord with $A_H = A_y = B \cdot X$ and $A_G = A_x = - B \cdot Y$. Hence our model describes also the cyclotron motion of electrons in QHE, in accord with the presented canonical quantization of the action (10) which is given among others by: $e [ \hat{A}_x  \ \ , \ \ \hat{X} ] = e B [ \hat{Y} \ \ , \ \ \hat{X} ] = - i \hbar$ as in the phenomenological cyclotron relation (4). Therefore the quantum model which is represented by relations (10)-(13) is an adequate model to describe the IQHE in the above discussed manner, since this model describes both the quantized electromagnetic interactions between electrons and the magnetic field in the cyclotron motion on the IQHE sample, as well as between Hall and gate potentials in IQHE.
Further note that since the uncertainty relations or inequalities (6), (7) or (13) gives a reciprocal relation between the general measurable values of the pair of the involved quantities. Therefore the minimum value of one of these quantities is accompanied by the maximum values of the other one. In other words the case where $(\Delta X_m)_{(minimum)} = l_B$, is accompanied by $(\Delta A_m)_{(maximum)} > \displaystyle{\frac{\hbar}{e l_B}}$. Hence the $( = )$ equality case within the $( \geq )$ inequality cases of the uncertainty relations is given for the case where both quantities have their minimum values, i. e.: for $(\Delta A_m)_{(minimum)} \cdot (\Delta X_m)_{(minimum)} = \displaystyle{\frac{\hbar}{e}}$. On the other hand the minimum value of $\Delta A_m$ is given by $(\Delta A_m)_{(minimum)} = B \cdot l_B$, in view of $A_m = \epsilon_{mn} B \cdot X_n$ or $\Delta A_m = |\epsilon_{mn}| B \cdot \Delta X_n$, since $(\Delta X_m)_{(minimum)} = l_B$. Thus for the equality case where $(\Delta A_m)_{(minimum)} = B \cdot l_B$ and $(\Delta X_m)_{(minimum)} = l_B$, one obtains $(B \cdot l_B ^2 = \displaystyle{\frac{\hbar}{e}})$ which is the definition of the magnetic length. In this sense we obtain the defintion of the magnetic length from the uncertainty relation (13) in accord with the canonical quantization of the two dimensional model. A fact which confirms again the consistency of this model.   

Moreover note that recent results for the Hall photovoltage imaging of the edge of quantum Hall device \cite{drop}, confirms also our uncertainty relation for the electromagnetic potential, if one considers $\Delta A_H$ as the scale of potential drops on the edge. Recall that on the one hand the quantized Hall potential is given in our model by $\Delta A_H \geq \displaystyle{\frac{\hbar}{e \Delta Y}}$ which means that: If the sample and experiment are prepared in a way that the position uncertainty $\Delta Y$ of electrons moving on the sample becomes large, then one should expect lower values for the quantized Hall voltage $\Delta A_H$. In other words: $(\Delta Y)_{(large)} \rightarrow (\Delta A_H)_{(low)}$. On the other hand the $(\Delta Y)_{(large)} > l_B$  value is given for the case of Ref. \cite{drop}, by $(\Delta Y)_{(large)} > 1.4 \times 10^{- 6} \ cm$, since the value of magnetic length is given by: $l_B \approx 1.4 \times 10^{- 6} \ cm$ for this case. Hence for a higher value of $\Delta Y$, i. e. for a value about $(\Delta Y)_{(large)} \approx  10^{- 5} \ cm$ which is choosen just greater than $l_B \approx 1.4 \times 10^{- 6} \ cm$ , one obtains $(\Delta A_H)_{(low)} \approx 10^{-3} \ cm$ for Hall voltage, in accord with our uncertainty relation \cite{h}. This is equal to the measured value of $10 \ \mu m$ for the length scale of Hall photovoltage imaging of the edge of quantum Hall devices, i. e. for the length scale of the edge confining potential, in the experiment reported in the Ref. \cite{drop}. Note that, as it is discussed above, the square of length $L^2$ is an invariant in such a two dimensional model which is defined on a two dimensional manifold $( \sim X_m \ m = 1, 2)$. This means that: $L^2 = (invariant) = (constant) \sim L^0$, since a constant has no dimension. Equivalently one has in this case also $L^1 \sim L^{-1}$. Therefore the scale of Hall voltage, which is of dimension $L^{-1}$, is given by a quantity of dimension $L$ in $cm$ or $\mu m$.

Further note that also early experiments on the Hall photovoltage imaging of the edge of quantum Hall devices confirm the uncertainty relation (15) and thereby this model \cite{alt}.

In conclusion note that in the two dimensional model we are able to calculate the quantized values of Hall conductivity or resistivity directly from the quantization of action function:
The above introduced canonical quantization or Bohr-Sommerfeld quantization postulate of the two dimensional single electron model of QHE, i. e. $e \oint\limits_{contour} A_m dx^m = N h$ can be rewritten for a real QHE sample with an electronic density of $n$ by $e \oint\limits_{contour} A_m dx^m = e \epsilon_{mn} \int \int\limits_{surface} \rho_H \cdot n e dx^m \wedge dx^n = N h$, in view of $A_m = B x^n \epsilon_{mn}$ and $(\sigma_H)^{-1} = \rho_H := \displaystyle{\frac{B}{ne}}$. Considering the constancy of $\rho_H$ and the total number of electrons on the sample $N^{\prime} := \int \int\limits_{surface} ne dx^m \wedge dx^n$, one obtains from the canonical quantization of the action (10) the following relation for Hall conductivity:

\begin{equation}
\sigma_H = \nu \displaystyle{\frac{e^2}{h}}  \ ,
\end{equation}

where the filling factor is defined by: $\nu = \displaystyle{\frac{N^{\prime}}{N}} ; \  (N, N^{\prime}) \in {\mathbf Z}$. Hence the model describes IQHE and FQHE for $\nu = \displaystyle{\frac{N^{\prime}}{N}} \in {\mathbf Z}$ and $\nu = \displaystyle{\frac{N^{\prime}}{N}} \not\in {\mathbf Z}$, respectively. One can prove the consistency of this definition of filling factor, i. e. its equivalence with the phenomenological definition: $\nu := 2 \pi n l_B ^2$, if one rewrites this one by $\nu := \displaystyle{\frac{n h}{e B}}$, in accord with $l_B ^2 = \displaystyle{\frac{\hbar}{e B}}$. Hence one obtains from $\nu := \displaystyle{\frac{n h}{e B}}$ the above definition of filling factor $\nu = \displaystyle{\frac{N^{\prime}}{N}}$, in view of $N^{\prime} = n \cdot a = \epsilon_{mn} \int \int dx^m \wedge dx^n$ and $e \epsilon_{mn} \int \int B dx^m \wedge dx^n = e B \cdot a = N h$ where $a$ is the area of sample.
The reason that FQHE appears on samples with smaller electronic density under stronger magnetic field, whereas IQHE appears on samples with higher electronic density under lower magnetic field, may lie in the discussed equivalence between the two definitions of filling factor: $\nu = \displaystyle{\frac{n h}{e B}} = \displaystyle{\frac{N^{\prime}}{N}}$. It is obvious from this relation that the FQHE case with smaller $(n)$ and the larger $(B)$ implies smaller filling factors than the IQHE case with larger $(n)$ and smaller $(B)$. Thus for a given sample area $(a)$, a higher magnetic field $B$ implies higher quantum numbers $N$.

Further note that, in view of the fact, that the model described here is a very general model of {\it two dimensional} quantum electrodynamics, therefore the presented uncertainty relations (15) are new uncertainty relations for the quantized electromagnetic field, i. e. the physical photon, with its two degrees of freedom only. In this sense, it is shown that just the quantum effects require a two dimensional quantum electrodynamics, since the four dimensional QED does not explain the magnetic quantum effects.

\bigskip

\unitlength 1cm
\begin{flushleft}

\begin{picture}(15,7.5)




\put(3.2,1){\epsfxsize=10cm \epsfbox{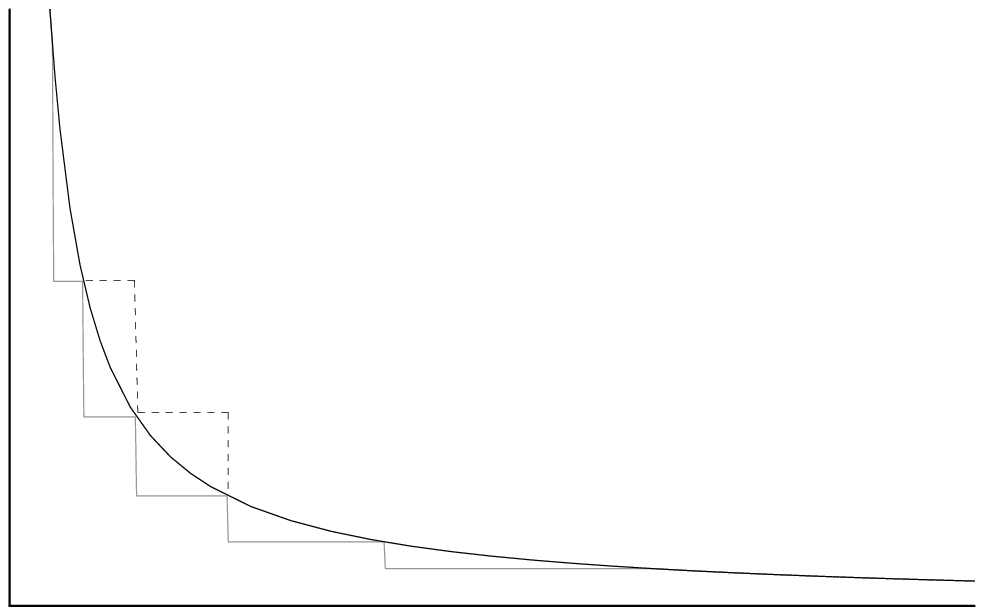}}
\put(4.3,2.18){\epsfysize=0.9cm \epsfbox{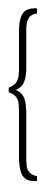}}
\put(4.53,1.97){\epsfxsize=1.05cm \epsfbox{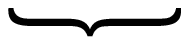}}
\put(3.2,7.3){$\displaystyle A_H$}
\put(13,0.9){$A_G$}
\put(6,5.5){$  \displaystyle a_1 = a_i = a_{i+1} = a_n = \Delta A_H \cdot \Delta A_G = \frac{\hbar B}{e}$}
\put(4.6,1.8){$ \displaystyle \Delta A_G$}
\put(3.55,2.6){$\displaystyle \Delta A_H$}
\put(4.65,2.55){$\displaystyle a_{i+1}$}
\put(4.2,3.4){$\displaystyle a_i$}
\end{picture}
\end{flushleft}

Fig. 1: The schematic classical curve $(A_H = A_y = \displaystyle{\frac{B}{A_G}} = \displaystyle{\frac{1}{Y}})$ and the schematic quantum curve $(\Delta A_H = \displaystyle{\frac{\hbar B}{e \Delta A_G}})$ are represented by the full line and the doted line, respectively. The example of parallelograms with equal areas are drawn on two of QHE plateaus. All such areas $a_i$ are equal.

\bigskip
Footnotes and references


\begin{thebibliography}{100}
\bibitem{Bsing}
We consider here in a first step the IQHE case, where in the absence of spin and the electron-electron interaction, the {\it single electron picture} is available. For multi-electron case one should replace the charge $e$ by $Q := N^{\prime} e$, where $N^{\prime}$ is the total number of electrons on the sample.
\bibitem{rel}
With potential, we mean always a gauge invariant potential difference: $\delta A_m := A_m - A_0$ where the background voltage $A_0$ can be rescalled to zero.
\bibitem{new}
Note that the operator $\hat{P}_m := - i \hbar \partial_m$ commute with the operator $\hat{A}_m :=  B \cdot \hat{X}^n \epsilon_{mn}$.
\bibitem{h}
We use for the $\displaystyle{\frac{\hbar}{e}} \approx 10^{-8} \ \displaystyle{\frac{erg . s}{Amper . s}}$ in accord with $h \approx 6.67 \times 10^{-27}$ and $e \approx 1.6 \times 10^{- 19} \ Amper . s$.
\end{thebibliography}
\end{document}